\documentclass{ws-procs975x65}
\newcommand{\be}{\begin{equation}}
\newcommand{\ee}{\end{equation}}
\newcommand{\bea}{\begin{eqnarray}}              
\newcommand{\eea}{\end{eqnarray}}

\begin{document}

\title{MG13 proceedings: 
  VILKOVISKY-DeWITT EFFECTIVE POTENTIAL REVISITED IN GAUGE-FREE FRAMEWORK}

\author{
  {SRIJIT BHATTACHARJEE}
}

\address{Astroparticle Physics \& Cosmology Division,\\
  ~Saha Institute of Nuclear Physics, India\\
  1/AF Bidhannagar, Kolkata-64\\
  E-mail: srijit.bhattacharjee@saha.ac.in}

\begin{abstract}
An alternative approach to scalar quantum electrodynamics has been
proposed where the usual gauge redundancy of the theory do not manifest.
The gauge-dependence of Coleman- Weinberg effective potential is resolved using gauge-free approach leading to a unique scalar to vector mass ratio. The gauge-free result is compared with the one obtained in Vilkovisky-DeWitt’s geometric framework and it is shown that it differs from the standard result reported earlier. It is also shown that for reparametrization invariance of the Coleman-Weinberg potential one must combine the Vilkovisky-DeWitt’s method with the gauge-free approach.
\end{abstract}

\keywords{gauge-invariance, gauge-free, Coleman-Weinberg potential,
Vilkovisky-DeWitt}

\bodymatter

\section{Introduction}
\label{sec:intro}
In gauge theories, the off-shell Green's functions in quantum field theory often offer ambiguous results due to gauge dependence. Coleman-Weinberg \cite{CW} effective potential being one of those off-shell quantities, also suffers from this problem \cite{Jackiw}. Moreover, there exists an additional ambiguity which spoils the uniqueness of effective potential (or effective action) due to field reparametrizations. Both of these issues can be handled with the help of Vilkovisky-DeWitt's (VD) geometric method to obtain a unique effective action \cite{vilk,DeWitt}. On the other hand, rewriting charged matter fields in polar coordinates enables a reformulation of gauge theories in which the dynamical variables are manifestly inert under gauge transformations. This {\it gauge-free} method essentially eschews the redundant degrees of freedom from the outset of a theory and would give a physical result \cite{pmb}. However, to obtain a reparametrization invariant effective potential using gauge-free approach one also 
requires to employ VD technique. 

In this article we give a brief outline of gauge-free framework and state the result of one-loop effective potential for gauge-free scalar QED in VD approach.
\section{Gauge-free prescription}
\label{sec:gaugefree}
In standard model, to describe any theory of charged fields we have to introduce gauge fields. All charged matter fields can be decomposed into a radial and a phase part ${\Phi} = {\phi} \exp i\theta$. The radial part remains inert under the gauge transformation but the phase part gets transformed under it(${\phi}^{\omega}\rightarrow {\phi}$ and $\theta^{\omega} \rightarrow  \theta\,+\omega$). We know that gauge transformation acts only on the unphysical degrees of freedom. We can decompose an abelian gauge field $A_{\mu}$ into a transverse(physical) and a longitudinal(unphysical) part. This decomposition can be easily demonstrated by a covariant projector in case of an abelian gauge field, ${\bf A}_{{\cal P}\mu}={\cal P}_{\mu \nu}A^{\nu}=(\eta_{\mu\nu}-\partial_{\mu}\Box^{-1}\partial_{\nu})A^{\nu}$. The physical gauge potential being transverse, satisfies $\partial \cdot  {\bf A}_{\cal P} =0$. The longitudinal part of gauge potential can be represented as $ A_{\mu}^L(x)=\partial_{\mu}\int
d^4 x^{'} G(x, x^{'} ) \partial^{'} · A(x^{'})=\partial_{\mu}a(x)$. It is clear that only the longitudinal part of gauge potential gets changed by an abelian gauge transformation like $A\rightarrow A~+~\partial e^{-1}\omega(x)$. Now in scalar QED we have the
phase field $\theta$, which appears in the action only
through its first order derivative $\partial \theta : S[\Phi] =
S[\phi, \partial\theta]$. The gauge-free prescription for coupling the
gauge-free vector potential ${\bf A}_{\cal P}$ to $\Phi$ is exceedingly simple
: leaving $\phi$ as it is in the action, simply replace $\partial \theta
\rightarrow \partial \theta - e {\bf A}_{\cal P}$, so that $S[\Phi] \rightarrow
S[\phi, \partial \theta - e {\bf A}_{\cal P}] + S_{free} [{\bf A}_{\cal P}]$.
The gauge-free action for scalar QED in $(\rho,\Theta)$ basis becomes 
\bea
S[\rho,\Theta,{\bf A}_{\cal P}] = \int d^4x \left[ \frac12 (\partial
\rho)^2 + \frac12 e^2 \rho^2 ({\bf A}_{\cal P} - e^{-1}\partial \Theta)^2
~- \frac12 (\partial {\bf A}_{\cal P})^2 -  {\lambda \over 4!}\rho^4 \right],\label{Lsqed}
\eea
 with $\Theta=\theta-eA_L$ is a gauge inert {\it physical} field because gauge transformations of $\theta$ and $A_L$ cancel each other. 
\section{Gauge-free VD Effective Potential for scalar QED}
\label{sec:gfVD}
The (VD) geometric effective potential in one-loop approximation is given by \cite{vilk, DeWitt}
\be
 V_{VD}^{(1)}[\Phi]=S(\Phi)+{\hbar \over i}~\ln\mu[\phi]+{\hbar \over 2i}Tr\ln[\nabla_m\nabla_nS(\Phi)]+O(\hbar^2) 
\ee
Where the configuration space is promoted to a manifold (for details see \cite{vilk}). The one-loop effective potential computed from (\ref{Lsqed}) in gauge-free VD framework is:
\bea
V^{gfVD}_{eff}(\rho_c) &=& {\lambda \over 4!} \rho_c^4 + {1 \over 64 \pi^2}\left(3e^4  + {5\lambda^2 \over 18} \right)\hbar \rho_c^4
\left[\log {\rho_c^2 \over M^2} - {25\over 6} \right]~.
\label{epvdgf}
\eea
Where $\rho_c$ is the spacetime independent saddle point of the $\rho$ field and $M$ is the renormalization scale. This result \cite{sbpm} differs from the earlier one reported by Kunstatter \cite{Kun}.


\begin{thebibliography}{99}
\bibitem {CW} S. Coleman and E. Weinberg, Phys. Rev. {\bf D 7} (1973) 1888.
\bibitem {Jackiw} R. Jackiw, Phys. Rev. {\bf D 9} (1974) 2276.
\bibitem{vilk} G. A. Vilkovisky, Nucl. Phys. B 234, 1984, 125; G. A. Vilkovisky, in {\it Quantum Theory of Gravity}, edited by S. Christensen (Hilger, Bristol, 1984).
\bibitem{DeWitt} B. S. DeWitt, in \textit{Architecture of Fundamental Interactions at Short Distances}, proceedings of the Les Houches
 Summer School, Les Houches, France, 1985, edited by P. Ramond and R. Stora (Les Houches Summer School Proceedings Vol. 44)(North Holland, Amsterdam, 1987).
\bibitem {pmb} S. Bhattacharjee and P. Majumdar, Phys. Rev. {\bf D83} 085019, 2011. arXiv: 1006.1712.
\bibitem{sbpm}S. Bhattacharjee and P. Majumdar (in preparation).
\bibitem {Kun} G. Kunstatter, in {\it Super Field Theories}, proceedings of the NATO Study Institute, Vancouver, Canada, 1986, edited
by H. C. Lee (NATO ASI Series B, Vol. 160)(Plenum, New York, 1987).
\end{thebibliography}
\end{document}